\documentclass[doublecol]{epl2} 
\usepackage{amsmath}
\usepackage{graphicx}

\newcommand{\sgn}{\mathop{\mathrm{sgn}}\nolimits}

\title{Dynamics of social balance under temporal interaction}
\shorttitle{Dynamics of social balance under temporal interaction} 
\author{Ryosuke Nishi\inst{1,2} \and Naoki Masuda\inst{3,4,5}\thanks{E-mail: \email{naoki.masuda@bristol.ac.uk}}}
\shortauthor{Ryosuke Nishi and Naoki Masuda}

\institute{                    
  \inst{1} National Institute of Informatics, 2-1-2 Hitotsubashi, Chiyoda-ku, Tokyo 101-8430, Japan\\
  \inst{2} JST, ERATO, Kawarabayashi Large Graph Project, 2-1-2 Hitotsubashi, Chiyoda-ku, Tokyo 101-8430, Japan\\
  \inst{3} Department of Engineering Mathematics, Merchant Venturers Building, University of Bristol,
Woodland Road, Clifton, Bristol BS8 1UB, United Kingdom\\
  \inst{4} CREST, JST, 4-1-8 Honcho, Kawaguchi-shi, Saitama 332-0012, Japan\\
  \inst{5} Department of Mathematical Informatics, The University of Tokyo, 7-3-1 Hongo, Bunkyo-ku, Tokyo 113-8656, Japan\\
}

\abstract{
Real social contacts are often intermittent such that a link between a pair of nodes in a social network is only temporarily used. Effects of such temporal networks on social dynamics have been investigated for several phenomenological models such as epidemic spreading, linear diffusion processes, and nonlinear oscillations. Here, we numerically investigate nonlinear social balance dynamics in such a situation. Social balance is a classical psychological theory, which dictates that a triad is balanced if the three agents are mutual friends or if the two of them are the friends of each other and hostile to the other agent. We show that the social balance dynamics is slowed down on the temporal complete graph as compared to the corresponding static complete graph.
}

\begin{document}

\maketitle

\section{Introduction}
Social contacts such as face-to-face meetings and communications via emails often occur intermittently. In terms of social networks, links connecting pairs of individuals are activated and deactivated over time because individuals may be mobile and engaged in different types of activity even on a short time scale. Networks with dynamic links are collectively called the temporal networks, and an increasing amount of network data containing temporal information about links is available~\cite{Holme2012,Holme2013}. Temporal patterns of links influence various properties of networks including dynamics on networks. For example, in epidemic spreading (see~\cite{Masuda2013b} for a review), linear diffusion~\cite{Starnini2012,Masuda2013a,Scholtes2013,Delvenne2013,Rocha2014}, nonlinear synchronous oscillations~\cite{Fujiwara2011}, consensus formation~\cite{Fernandez-Gracia2011,Baxter2011,Takaguchi2011}, and the so-called naming game~\cite{Maity2012}, dynamics occurring on temporal networks can substantially differ from those occurring on the corresponding aggregate (i.e., static) networks. We consider that understanding yet other social dynamics models on temporal networks would strengthen understanding of temporal networks.

A link connecting two nodes in a social network represents a friendly, antagonistic, or different type of relationship depending on the definition of link. In particular, friendly and antagonistic links would coexist in a single social network. The concept of the social balance is a classical theory founded in social psychology dating back to Heider~\cite{heider1946attitudes, Wasserman1994Social}. It is a framework for classifying triads composed of friendly or antagonistic links into stable and unstable relationships, as shown in Fig.~\ref{fig:def_balance}. Heider's social balance for triads was generalized to a concept of balance for networks, i.e., the so-called structural balance~\cite{cartwright1956structural}. Heider's social balance and its variants are empirically found in human~\cite{Harary1961,Healy1973,Moore1978,Doreian1996,Doreian1996a,Leskovec2010,Szell2010,Szell2010a,Facchetti2011} and other animal~\cite{Ilany2013} societies (also see~\cite{Kovanen2011,Gallos2012} for related measurements).

Early studies of social balance focused on static properties (e.g., measurement of social balance in a given network with signed links).
Dynamics of Heider's social balance is a relatively recent research topic. Antal and colleagues proposed dynamical models with binary signed links (i.e., positive or negative links) and a well-mixed population (i.e., complete graph)~\cite{Antal2005,Antal2006b}. Unbalanced final states may be reached in their models, and the organization of unbalanced states was later studied in terms of the energy landscape~\cite{Marvel2009}. According to numerical simulations of a model with continuous-valued links, the global social balance was eventually reached in the complete graph~\cite{Gawronski2005,KUAKOWSKI2005}, which was theoretically proved in later work~\cite{Marvel2011}. In fact, the probability that the balanced state is reached in finite time tends to unity as the number of agents tends to infinity. In addition, they derived the time to reach balanced states~\cite{Marvel2011}. Effects of asymmetric initial conditions and asymmetric networks were also explored~\cite{Traag2013}. The possibility that a given agent locally perturbs the link weight to secure a desired global balance state has also been examined~\cite{Summers2013}. The same dynamics in the two-dimensional triangular lattice was examined in Ref.~\cite{Radicchi2007}. 

These dynamical models of social balance were investigated in static networks. In this letter, we aim to clarify the effect of temporal interaction on the speed of the dynamics of social balance. We do so by numerically investigating a model that extends those proposed in Refs.~\cite{KUAKOWSKI2005, Marvel2011} to the case of temporal interaction. We model the temporal interaction by assuming that only one link in the aggregate network is active at a time for a fixed duration.

\section{Model}
\begin{figure}[t]
\centerline{
\includegraphics[width=0.9\hsize]{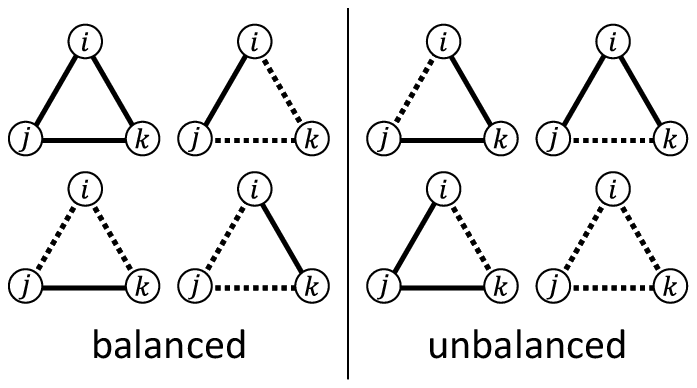}
}
\caption{The balanced and unbalanced states of triad $\left\{i, j, k\right\}$. Solid and dashed links represent positive and negative links, respectively.}
\label{fig:def_balance}
\end{figure}

Consider the undirected complete graph composed of $N$ nodes labeled $1,\ldots,N$. We denote by $x_{ij}$ ($=x_{ji}$; $1 \le i, j \le N$) the weight of the link between nodes $i$ and $j$, which may be negatively valued. Positive and negative values of $x_{ij}$ represent a friendly and hostile relationship between $i$ and $j$, respectively. Because of the underlying complete graph, any triplet of nodes $i$, $j$, and $k$ ($1 \le i, j, k \le N$), where $i$, $j$, and $k$ are mutually different, forms a connected triad. A triad $\left\{i, j, k\right\}$ is defined as balanced if $x_{ij} x_{jk} x_{ki} > 0$. The balanced triads are enumerated in Fig.~\ref{fig:def_balance}. In a balanced triad, the three nodes are pairwise friends, or two nodes are friends of each other and in conflict with the third node. Triad $\left\{i, j, k\right\}$ is defined as unbalanced if $x_{ij} x_{jk} x_{ki} < 0$, representing the frustrating relationship among the three nodes, as shown in Fig.~\ref{fig:def_balance}. We define $N_{\Delta}$ as the number of triads, i.e., $N_{\Delta} = N(N-1)(N-2)/6$. The population of the $N$ nodes is defined as balanced if all $N_{\Delta}$ triads are balanced. Otherwise, the population is unbalanced.

Ku{\l}akowski and colleagues proposed a continuous-time dynamics given by
\begin{align}
\dfrac{{\rm d}x_{ij}}{{\rm d} t} = \sum_{k=1, k\neq i,j}^N x_{ik} x_{kj},
\label{eq:dxijdt_aggr_previous_S0}
\end{align}
where $1 \le i \neq j \le N$, and $t$ denotes the time~\cite{KUAKOWSKI2005}. In Eq.~(\ref{eq:dxijdt_aggr_previous_S0}), the dynamics of $x_{ij}$ is driven by a psychological force exerted on each triad $\left\{i, j, k\right\}$ ($k=1,\ldots,N, k \neq i,j$). This force pushes $x_{ij}$ toward the balanced state. The balanced state of the population once reached is maintained forever, rendering it a stable equilibrium of the dynamics given by Eq.~(\ref{eq:dxijdt_aggr_previous_S0}). 

In agent-based simulations of the dynamics given by Eq.~(\ref{eq:dxijdt_aggr_previous_S0}), some $x_{ij}$ diverges much faster than others to make numerical simulations difficult. To avoid this difficulty, we restrict $x_{ij}$ to the range $[-R, R]$ for some $R>0$ by multiplying $(1 - x_{ij}^2/R^2)$ to the right-hand side of Eq.~(\ref{eq:dxijdt_aggr_previous_S0})~\cite{KUAKOWSKI2005}. In the numerical simulations carried out in Ref.~\cite{KUAKOWSKI2005}, the balanced state was reached with $R=5$. Furthermore, we normalise the right-hand side of Eq.~(\ref{eq:dxijdt_aggr_previous_S0}) by dividing it by the number of summands, i.e., $N-2$, to obtain   
\begin{align}
\dfrac{{\rm d}x_{ij}}{{\rm d} t} = \dfrac{1}{N-2} \left(1 - \dfrac{x_{ij}^2}{R^2}\right) \sum_{k=1,k \neq i,j}^N x_{ik} x_{kj}. 
\label{eq:dxijdt_aggr_modefied_S0}
\end{align}
It should be noted that the dynamics given by Eq.~(\ref{eq:dxijdt_aggr_modefied_S0}) preserves the symmetry $x_{ij}=x_{ji}$ for any $i$ and $j$ if the same symmetry is initially satisfied for all $i$ and $j$.

Our main concern is to investigate the dynamics of social balance on the temporal complete network. We define the dynamics as follows. First, we select a pair of nodes $i_r$ and $j_r$ with the equal probability from the $N$ nodes, and let $x_{i_rj_r}$ evolve according to Eq.~(\ref{eq:dxijdt_aggr_modefied_S0}) for time $\tau$ without changing the other $x_{ij}$'s. Then, we pick a pair of nodes in the same random manner, independently of the previous choice of the node pair, and apply Eq.~(\ref{eq:dxijdt_aggr_modefied_S0}) exclusively for the selected pair for time $\tau$. We repeat this procedure. 

The aforementioned update scheme is equivalent to the edge sequences with replacement examined in Ref.~\cite{Masuda2013a}. The dynamics approaches the aggregate dynamics, i.e., the one on the aggregate network in which all links are simultaneously used (Eq.~(\ref{eq:dxijdt_aggr_modefied_S0})) in the limit $\tau \to 0$. A large $\tau$ value represents strong temporality of the dynamics.

\section{Results}
In the following numerical simulations, we judge the state of each triad $\left\{i,j,k\right\}$ as follows. We define $\sgn x_{ij}=1$ if $x_{ij} \ge \epsilon$, $\sgn x_{ij}=-1$ if $x_{ij} \le -\epsilon$ and $\sgn x_{ij}=0$ if $|x_{ij}| < \epsilon$, where $\epsilon=1.0 \times 10^{-6}$. Triad $\left\{i,j,k\right\}$ is regarded to be balanced if and only if $\sgn x_{ij} \sgn x_{jk} \sgn x_{ki}=1$. 
Once the population reaches a balanced state, all $x_{ij}$'s approach $\pm R$ without changing the sign. Therefore, the value of $\epsilon$ does not affect the following results. We have chosen a small $\epsilon$ value to accelerate numerical simulations and avoid possible effects of the rounding.

We set $N=200$ and $R=10$ unless otherwise stated. The initial value of each $x_{ij}$ ($=x_{ji}$) for any $1 \le i \neq j \le N$ obeys the independent Gaussian distribution with mean $\mu$ and standard deviation $1$. In each run, we measure the number of unbalanced triads, denoted by $N_{\Delta}^{\rm U}$, at $t=0, 10, 20, \ldots$. We stop the run at $t=T$ when all triads become balanced for the first time. If $t$ reaches $2 \times 10^6$ without realizing the balance at the population level, we terminate the run and regard it as an unfinished run. For a given parameter set, we repeat $10^3$ runs starting from different initial conditions. By varying $\tau$, we bridge the aggregate dynamics (small $\tau$) and temporal dynamics (large $\tau$). We use the same $10^3$ initial conditions for different $\tau$ values. Unless otherwise stated, the following results are averages over the runs that have finished for all the eight $\tau$ values (i.e., $\tau = 0.01, 0.02, 0.05, 0.1, 0.22, 0.5, 1, 2.25$). In other words, we discard the initial conditions for which the run does not terminate for at least one $\tau$ value. In this way, we exclude the possibility that different initial conditions cause the dependence of the following numerical results on $\tau$. At most $\approx 8$\% of the runs are discarded (Table~\ref{tab:N_DiscRuns}). The number of discarded runs was large for small $\mu$. The initial conditions for different $\mu$ values are generally different.

\begin{table}
\caption{Number of discarded runs among the $10^3$ runs. ``with replacement" and ``without replacement" are two implementations of the temporal dynamics.}
\label{tab:N_DiscRuns}
\begin{center}
\small
\begin{tabular}{lccc}
\hline
\multicolumn{1}{c}{model} & $\mu=-1$ & $\mu=0$ & $\mu=1$ \\ \hline
Eq.~(\ref{eq:dxijdt_aggr_modefied_S0}), with replacement    & 37 & 28 & 0 \\
Eq.~(\ref{eq:dxijdt_aggr_modefied_S0}), without replacement & 35 & 22 & 0 \\
Eq.~(\ref{eq:dxijdt_aggr_modefied_S1}), with replacement    & 79 & 49 & 0 \\
Eq.~(\ref{eq:dxijdt_aggr_modefied_S1}), without replacement & 75 & 29 & 0 \\
\hline
\end{tabular}
\end{center}
\end{table}

\begin{figure*}[htbp]
\centerline{
\includegraphics[width=\hsize]{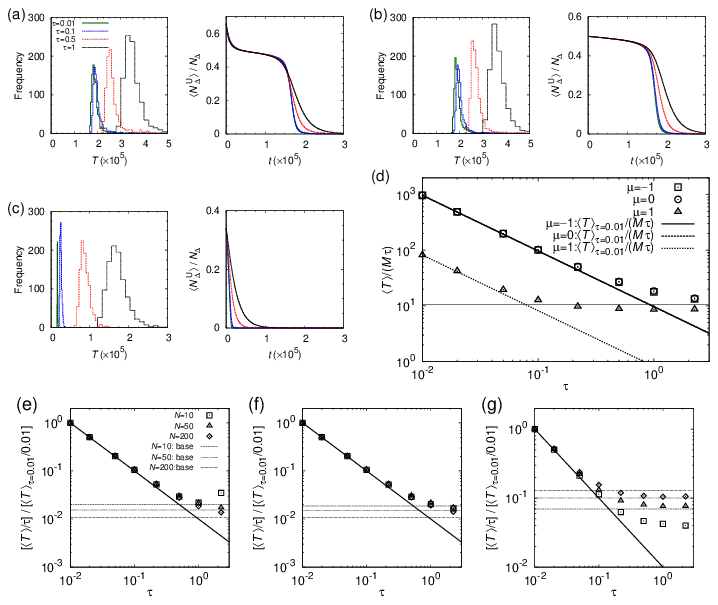}
}
\caption{
Behavior of the dynamics given by Eq.~(\ref{eq:dxijdt_aggr_modefied_S0}) under edge sequences with replacement. (a)--(c) Distributions of $T$ on the basis of $10^3$ runs (left panels). (a) $\mu=-1$. (b) $\mu=0$. (c) $\mu=1$. Time courses of the ensemble average of $N_{\Delta}^{\rm U} / N_{\Delta}$ on the basis of the same $10^3$ runs are shown in the right panels of (a), (b), and (c). (d) Number of updating events before the balance is reached, $\left\langle T \right\rangle/(M \tau)$, plotted against $\tau$ for the results shown in (a), (b), and (c). We also show $\left\langle T \right\rangle_{\tau=0.01}/(M \tau)$ as guides to the eye (lines). The lines for $\mu=-1$ and $\mu=0$ almost overlap with each other.
(e)--(g) Relationship between the number of update events before the balance is reached and the population size $N$. The former quantity is normalized by the value at $\tau=0.01$. (e) $\mu=-1$. (f) $\mu=0$. (g) $\mu=1$. 
In (e)--(g), we also show a solid thick line corresponding to $0.01/\tau$ as guides to the eye. In (d)--(g), the thin (horizontal) lines correspond to the expected number of updates per link at which each link is selected at least once.}
\label{fig:SelfExci=0Rpl=1}
\end{figure*}

\begin{figure*}[htbp]
\centerline{
\includegraphics[width=\hsize]{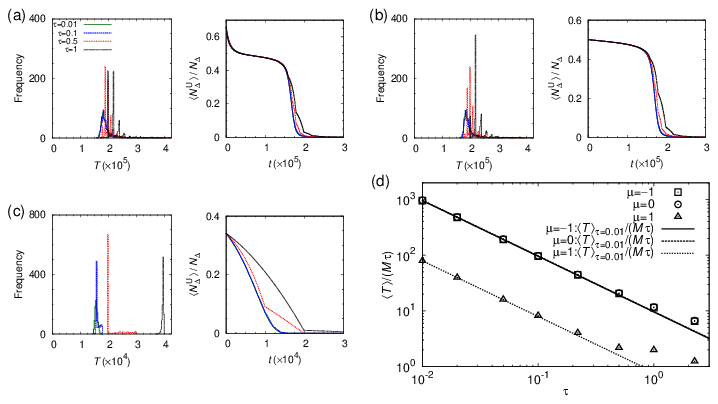}
}
\caption{ 
Results of the dynamics given by Eq.~(\ref{eq:dxijdt_aggr_modefied_S0}) under edge sequences without replacement. (a) $\mu=-1$. (b) $\mu=0$. (c) $\mu=1$. In (d), the lines for $\mu=-1$ and $\mu=0$ almost overlap with each other. See the caption of Fig.~\ref{fig:SelfExci=0Rpl=1} for details.
}
\label{fig:SelfExci=0Rpl=0}
\end{figure*}

Numerically obtained distributions of $T$ for $\mu=-1$, $\mu=0$, and $\mu=1$ are shown in the left panels of Figs.~\ref{fig:SelfExci=0Rpl=1}(a), (b), and (c), respectively. The figure shows that $T$ increases with $\tau$ for any $\mu$ on average. Therefore, the dynamics of structual balance slows down on temporal complete graph as compared to aggregate complete graph. We remark that the distribution of $T$ is broad for large $\tau$. 

The ensemble average of the number of unbalanced triads divided by the number of triads, $\left\langle N_{\Delta}^{\rm U} \right\rangle / N_{\Delta}$, is shown for $\mu=-1$ and $\mu=0$ in the right panels of Figs.~\ref{fig:SelfExci=0Rpl=1}(a) and (b), respectively. In this measurement, we have not excluded the unfinished runs from the statistics. For different $\tau$ values, $\left\langle N_{\Delta}^{\rm U} \right\rangle / N_{\Delta}$ decays with similar time courses in an early stage of the dynamics. In a late stage of the dynamics, $\left\langle N_{\Delta}^{\rm U} \right\rangle / N_{\Delta}$ decays more slowly on temporal dynamics (i.e., large $\tau$) than aggregate dynamics (i.e., small $\tau$). In contrast, the decay rate of $\left\langle N_{\Delta}^{\rm U} \right\rangle / N_{\Delta}$ depends on $\tau$ already in an early stage when $\mu=1$ (right panel of Fig.~\ref{fig:SelfExci=0Rpl=1}(c)). This is because, for $\mu=1$, the initial $\left\langle N_{\Delta}^{\rm U} \right\rangle / N_{\Delta}$ value is smaller than that for $\mu=-1$ and $\mu=0$. The former value is in fact smaller than the value of $\left\langle N_{\Delta}^{\rm U} \right\rangle / N_{\Delta}$ below which the dependence of $\left\langle N_{\Delta}^{\rm U} \right\rangle / N_{\Delta}$ on $\tau$ emerges in the case of $\mu=-1$ and $\mu=0$ (i.e., $\left\langle N_{\Delta}^{\rm U} \right\rangle / N_{\Delta} \approx 0.45$).

To be more quantitative, we measure the number of update events before a balanced state is reached. An update event is defined as the application of Eq.~(\ref{eq:dxijdt_aggr_modefied_S0}) for time $\tau$ to a single link. The average number of updates per link before the balanced state is reached is given by $\left\langle T \right\rangle /(M \tau)$, where $\left\langle T \right\rangle$ is the time before the balanced state is reached, averaged over all runs that terminate for all the $\tau$ values. $M=N(N-1)/2$ represents the number of links.
We present $\left\langle T \right\rangle /(M \tau)$ instead of $\left\langle T \right\rangle$ because we can detect pathological cases with the former quantity. If just a single update event is sufficient to make the sign of the link consistent with the balanced state of the population, $\left<T\right>$ would be the time at which all links are selected at least once for the first time. If this is the case, $\left<T\right>$ increases with $\tau$; actually, $\left<T\right>$ would be proportional to $\tau$. We are not concerned with this trivial effect. Therefore, in the following, we request that $\left\langle T \right\rangle /(M \tau)$ is not too small.

The values of $\left\langle T \right\rangle /(M \tau)$ for different $\tau$ and $\mu$ values are shown in Fig.~\ref{fig:SelfExci=0Rpl=1}(d). In the figure, we also draw the lines $\left\langle T \right\rangle_{\tau=0.01} /(M \tau)$ for each $\mu$. If the numerical results are located on this line, $\left\langle T \right\rangle$ is independent of $\tau$. Figure~\ref{fig:SelfExci=0Rpl=1}(d) indicates that $\left\langle T \right\rangle /(M \tau)$ is larger than $\left\langle T \right\rangle_{\tau=0.01} /(M \tau)$ and that the deviation increases with $\tau$.
The thin solid (horizontal) line in Fig.~\ref{fig:SelfExci=0Rpl=1}(d) shows the expected number of updates per link at which each link is selected at least once. If the actual $\left\langle T \right\rangle /(M \tau)$ value is located on or below this line, the balance is presumably reached when the last link is used for the first time. Then, the slowing down of the dynamics is trivial, as explained above. Figure~\ref{fig:SelfExci=0Rpl=1}(d) indicates that the trivial effect comes into play when $\tau$ is larger than $\approx$ 2 for $\mu=-1$ and 0, and $\tau$ is larger than $\approx 0.1$ for $\mu=1$. Otherwise, the dynamics
slows down owing to temporal interaction, although the amount of deceleration is not large.

The dependence of $\left<T\right>$ on $N$ is shown in Figs.~\ref{fig:SelfExci=0Rpl=1}(e), (f), and (g) for $\mu=-1$, 0, and 1, respectively. In these figures, the number of update events normalized by the value for $\tau=0.01$, which approximates the aggregate dynamics, is plotted for making the comparison across different $N$ values easier. As in the case of Fig.~\ref{fig:SelfExci=0Rpl=1}(d), the plotted value
should be larger than the normalized expected number of updates at which each link is selected at least once (shown by horizontal lines in Figs.~\ref{fig:SelfExci=0Rpl=1}(e), (f), and (g)) to be able to say that the dynamics is slowed down
not just due to the trivial effect. The figures suggest that, for intermediate values of $\tau$, the amount of the slowing-down effect depends little on $N$ although the effect is generally small (for example, $\tau=0.22$ and 0.5 in Fig.~\ref{fig:SelfExci=0Rpl=1}(e)). For large $\tau$ values, the slowing-down effect diminishes as $N$ grows, but the effect is artificial anyways (i.e., close to the horizontal lines). 
Therefore, we expect that the main results (i.e., slowing down) persist in large populations.

Next, we investigate effects of the update scheme by using the so-called edge sequences without replacement~\cite{Masuda2013a}. To implement this update scheme, we first generate a random permulation on $M$ links. The links are used according to the order determined by the permutation. We denote the permutation by $P=\left(\left\{i_1,j_1\right\},\ldots,\left\{i_{M},j_{M}\right\}\right)$, where $i_{\ell} \neq j_{\ell}$ for $\ell=1,\ldots,M$. We apply Eq.~(\ref{eq:dxijdt_aggr_modefied_S0}) to $x_{i_1j_1}$ without changing the other $x_{ij}$'s for $0 \le t < \tau$. Then, we apply Eq.~(\ref{eq:dxijdt_aggr_modefied_S0}) to $x_{i_{\ell}j_{\ell}}$ for $(\ell-1)\tau \le t < \ell \tau$, where $\ell$ increases from $2$ through $M$. At $t=M\tau$, we redraw a random permutation $P$. Then, we let each $x_{i_{\ell}j_{\ell}}$ ($\ell=1,\ldots,M$) evolve for time $\tau$ according to the order determined by the renewed $P$. At $t = 2 M \tau$, we draw $P$ again. We repeat this procedure. 

Numerical results of the dynamics given by Eq.~(\ref{eq:dxijdt_aggr_modefied_S0}) under edge sequences without replacement are shown in Fig.~\ref{fig:SelfExci=0Rpl=0}. Similarly to the results shown in Fig.~\ref{fig:SelfExci=0Rpl=1}, $T$ tends to be large for large $\tau$. However, dependence of $T$ on $\tau$ is not as strong as in the case of the edge sequences with replacement (Fig.~\ref{fig:SelfExci=0Rpl=1}). 

In Fig.~\ref{fig:SelfExci=0Rpl=0}, the distributions of $T$ for large $\tau$ (i.e., $\tau=0.5$ and $1$) are rather discrete. Discreteness of the distributions of $T$ is particularly evident for $\tau\in\left\{0.5,1\right\}$. When $\mu=1$, the peaks are approximately located at $4\tau\times10^4 \approx 2 M \tau$, i.e, when each link has been applied exactly twice. When $\mu=0$ and $\mu=-1$, the peaks are approximately located at $20 M \tau$ for $\tau=0.5$ and $10 M \tau$ for $\tau=1$. Moreover, $\left\langle N_{\Delta}^{\rm U} \right\rangle / N_{\Delta}$ plotted against $t$ has discontinuous decay rates (right panels of Fig.~\ref{fig:SelfExci=0Rpl=0}(a)--(c)). We do not completely interpret these results as slowing down owing to temporal interaction; the slowing down at least partly owes to the discreteness originating from the update scheme. 

To assess the robustness of the results shown in Figs.~\ref{fig:SelfExci=0Rpl=1} and~\ref{fig:SelfExci=0Rpl=0}, we simulate a variant of the model in which self-loops are allowed~\cite{Marvel2011}. The modified model is given by 
\begin{align}
\dfrac{{\rm d}x_{ij}}{{\rm d} t} = \dfrac{1}{N} \left(1 - \dfrac{x_{ij}^2}{R^2}\right) \sum_{k=1}^N x_{ik} x_{kj}. 
\label{eq:dxijdt_aggr_modefied_S1}
\end{align}
It should be noted that $i=j$, $k=i$, and $k=j$ are allowed in Eq.~(\ref{eq:dxijdt_aggr_modefied_S1}), which contrasts to Eq.~(\ref{eq:dxijdt_aggr_modefied_S0}). Once the population is balanced and $x_{ii} \ge 0$ ($1 \le i \le N$), the balance will be kept from then on. Therefore, we decided to stop a run when all triads became balanced and $x_{ii} \ge \epsilon$ ($1 \le i \le N$) for the first time. Numerical results of the dynamics given by Eq.~(\ref{eq:dxijdt_aggr_modefied_S1}) are in fact indistinguishable from the results obtained from Eq.~(\ref{eq:dxijdt_aggr_modefied_S0}) for both edge sequences with replacement and those without replacement 
except for the number of discarded runs (Table~\ref{tab:N_DiscRuns}). The number of discarded runs obtained from Eq.~(\ref{eq:dxijdt_aggr_modefied_S1}) was larger than that obtained from Eq.~(\ref{eq:dxijdt_aggr_modefied_S0}).

\section{Discussion}
We numerically showed that the time to the global social balance is larger on the temporal than aggregate complete graph. The main result was robust against some model variations such as the inclusion of self-loops, different update schemes, and different values of $\mu$, i.e., parameter controlling the level of social balance in the initial condition.

We modeled temporal dynamics by sequential applications of randomly selected links. By doing so, we have implicitly neglected the effect of long-tailed distributions and temporal correlation of the interevent time, which are found in various types of temporal network data~\cite{Holme2012}. Therefore, the slowing down revealed in the present study is not derived from these features of temporal network data but from the non-simultaneity of link usage. In this sense, the present results are similar to the previous ones~\cite{Fujiwara2011, Masuda2013a, Scholtes2013}, which did not explicitly use the interevent time to model diffusive or ordering dynamics on temporal networks. Investigating other nonlinear dynamics under the matrix or edge sequence representation of temporal networks warrants future studies.

We confined ourselves to the undirected complete graph to focus on the effects of temporality. Previous studies on dynamics of social balance also assumed the undirected complete graph~\cite{Antal2005, Antal2006b, Gawronski2005, KUAKOWSKI2005}. Empirically, the balanced states for triads and larger networks have been mostly examined as static features~\cite{Harary1961,Healy1973,Moore1978,Doreian1996,Doreian1996a,Leskovec2010,Facchetti2011,Ilany2013} (but see~\cite{Szell2010,Szell2010a}). Generalisation of the current results to complex networks warrants future work.

By definition, a triangle with link weights 0.05, 0.05, and 0.95 is regarded to be balanced. Although the two links with weight 0.05 are close to neutrality,
the dynamics would enhance the two links if this triangle were isolated. A triangle with link weights $-0.05$, 0.05, and 0.05 is regarded to be unbalanced. Although all links are close to neutrality, the three link weights would diminish toward zero. These behaviors, which are inherent in the current and previous continuous-valued models, may not be so realistic. Different dynamical rules for continuous-valued links may make the model behave differently from the present model.




\end{document}